\documentclass[twocolumn,amssymb,preprintnumbers,superscriptaddress]{revtex4}
\usepackage{graphicx}

\begin{document}

\newcommand{\R}{Ce$_{1-x}$R$_x$CoIn$_5$}
\newcommand{\Y}{Ce$_{1-x}$Y$_x$CoIn$_5$}
\newcommand{\La}{Ce$_{1-x}$La$_x$CoIn$_5$}
\newcommand{\Gd}{Ce$_{1-x}$Gd$_x$CoIn$_5$}
\newcommand{\Yb}{Ce$_{1-x}$Yb$_x$CoIn$_5$}
\newcommand{\Nd}{Ce$_{1-x}$Nd$_x$CoIn$_5$}
\newcommand{\Sn}{CeCoIn$_{5-x}$Sn$_x$}
\newcommand{\Co}{CeCoIn$_5$}
\newcommand{\Rh}{CeRhIn$_5$}
\newcommand{\RhCo}{CeRh$_{1-y}$Co$_y$In$_5$}
\newcommand{\Ir}{CeIrIn$_5$}
\newcommand{\Tcoh}{$T_{\rm coh}$}
\newcommand{\Tc}{$T_c$}
\newcommand{\ie}{{\it i.e.}}
\newcommand{\eg}{{\it e.g.}}
\newcommand{\etal}{{\it et al.}}

\title{Incoherent non-Fermi liquid scattering in a Kondo lattice}

\author{Johnpierre~Paglione}
\email{paglione@umd.edu}
\author{T.~A.~Sayles}
\author{P.-C.~Ho}
\author{J.~R.~Jeffries}
\author{M.~B.~Maple}
\affiliation{Department of Physics, University of California at San Diego, La Jolla, California 92093, USA}


\maketitle

{\bf
One of the most notorious non-Fermi liquid properties of both archetypal heavy-fermion systems \cite{Seaman,Lohneysen,Trovarelli,Holmes} and the high-\Tc\ copper oxide superconductors \cite{Martin} is an electrical resistivity that evolves linearly with temperature, $T$. In the heavy-fermion superconductor \Co\ \cite{Petrovic_Co}, this linear behaviour was one of the first indications of the presence of a zero-temperature instability, or quantum critical point. Here, we report the observation of a unique control parameter of $T$-linear scattering in \Co, found through systematic chemical substitutions of both magnetic and non-magnetic rare-earth, R, ions into the Ce sub-lattice. We find that the evolution of inelastic scattering in \R\ is strongly dependent on the $f$-electron configuration of the R ion, whereas two other key properties -- Cooper-pair breaking and Kondo-lattice coherence -- are not. Thus, $T$-linear resistivity in \Co\ is intimately related to the nature of incoherent scattering centers in the Kondo lattice, which provides insight into the anomalous scattering rate synonymous with quantum criticality \cite{Coleman}.
}

Although recent theories \cite{Varma,Cox,Rosch,Holmes} provide possible routes to an explanation of $T$-linear resistivity -- found in both $f$-electron systems (\eg\ Y$_{1-x}$U$_x$Pd$_3$ \cite{Seaman}, CeCu$_{6-x}$Au$_{x}$ \cite{Lohneysen}, YbRh$_2$Si$_2$ \cite{Trovarelli}, CeCu$_2$Si$_2$ \cite{Holmes}), and the normal state of the cuprate superconductors \cite{Martin} -- a general interpretation awaits arrival \cite{Coleman}. Several paradoxical features regarding this anomalous scattering rate continue to defy understanding, such as its persistence over decades of energy scales \cite{Seaman,Trovarelli,Martin} and down to millikelvin temperatures in three-dimensional materials \cite{Seaman,Lohneysen,Trovarelli,Holmes,Petrovic_Co}, its coexistence with conventional ($T^2$) Hall angle scattering \cite{Mackenzie,Nakajima} and its inconsistency with one-parameter scaling \cite{Phillips}. Most recently, its observation over three decades of $T$ at the field-tuned quantum critical point (QCP) of \Co\ has been linked to a violation of the Wiedemann-Franz law \cite{Tanatar_WF}, an indication that this scattering rate is associated with the failure of Fermi-liquid theory in its most basic form.

Here we present a rigorous study of the effects of rare-earth substitution on three closely related features of the exotic metal \Co: unconventional superconductivity, Kondo lattice coherence and anomalous charge-carrier scattering. By diluting the Ce lattice of high-quality single-crystal specimens of \R\ with both non-magnetic (full or empty $4f$-shell) and stable-$4f$-moment substituent ions of varying size and electronic configuration, we are able to inject both ``Kondo holes'' (isoelectronic ions without magnetic moments) and strongly localized magnetic moments into the coherent Kondo lattice. This has allowed us to probe the spin exchange between the Ce$^{3+}$ localized magnetic moments and the spins of the conduction electrons involved in Cooper pairing, Kondo screening and anomalous transport in a controlled way, revealing a surprising contrast between the response of coherent phenomena and non-Fermi liquid behaviour to this perturbation.

\begin{figure}
 \centering
 \includegraphics[totalheight=2.6in]{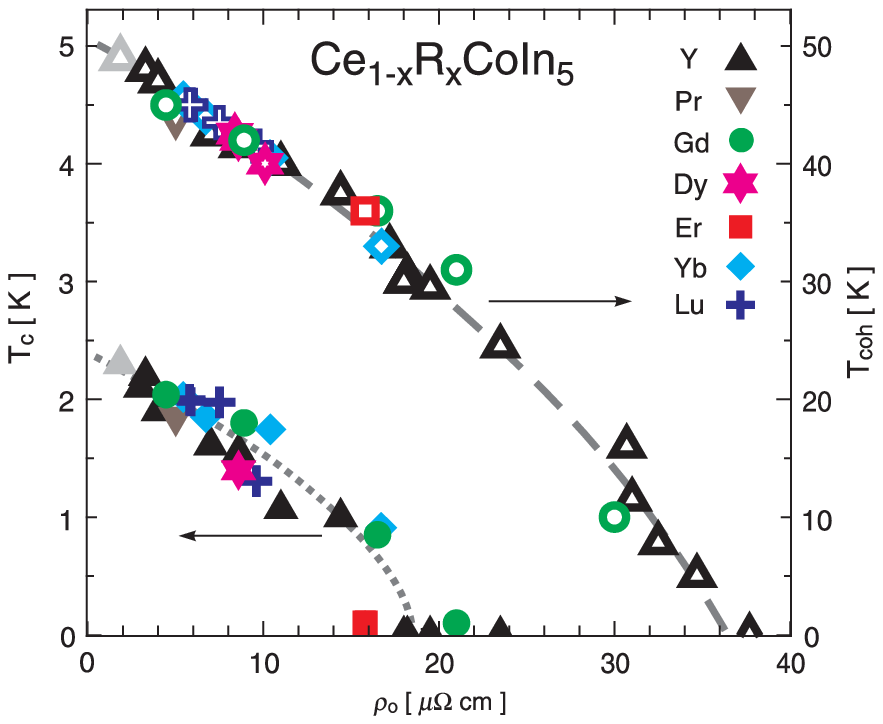}
 \caption{\label{fig:phase} 
 {\bf Dependence of superconducting transition temperature, $T_c$, and Kondo-lattice coherence temperature, \Tcoh\ of \R\ on rare-earth concentration.}
 Plotted as a function of residual resistivity ($\rho_0 \sim x$ -- see Methods section), this figure highlights the absence of any effect of the electronic configuration of replacement ion R on either \Tc\ (filled symbols) or \Tcoh\ (open symbols) as they evolve from $x=0$ (grey triangles). Note the lack of contrast between two particular species which are similar in all respects except $f$-electron filling: both Y$^{3+}$ ($5s^2 4d^1$) and Gd$^{3+}$ ($6s^2 4f^7 5d^1$) are isovalent with Ce$^{3+}$ ($6s^2 4f^1 5d^1$) and have nearly identical metallic radii of $1.801~\AA$, slightly smaller than that of Ce ($1.825~\AA$) and yielding a similarly small ($\sim 1 \%$) change of the lattice parameters upon substitution. The absence of an $f$-electron shell in Y$^{3+}$ leaves it non-magnetic, while the half-filled $f$-shell of Gd$^{3+}$ has the simplest configuration of the rare earths: a spherically symmetric $f$-shell with no orbital component ($J$=$S$=$7/2$, $L$=$0$) produces a large effective moment $\mu_{\rm eff}^2 = g^2J(J+1) = (7.9~\mu_B)^2$ with minimal effects from crystalline electric field anisotropy and spin-orbit coupling. The trends in $T_c$ and \Tcoh\ are also consistent with those found for the \La\ series \cite{Petrovic_La}.}
\end{figure}

Fig.~\ref{fig:phase} shows the evolution of both the superconducting transition temperature \Tc\ (identified by the transition in resistivity, $\rho$) and Kondo lattice coherence temperature \Tcoh\ (identified by the maximum in $\rho(T)$) for all rare earth substitutions made in \R\ through the complete range of concentrations where both features exist. As shown, the salient features are the same for all variants: as a function of residual resistivity ($\rho_0 \sim x$ -- see Methods), both \Tc\ and \Tcoh\ are suppressed to zero temperature at rates irrespective of the nature of the rare earth ion, which spans both magnetic (Pr$^{3+}$, Gd$^{3+}$, Dy$^{3+}$, Er$^{3+}$) and non-magnetic (Y$^{3+}$, Yb$^{2+}$, Lu$^{3+}$) $f$-electron configurations. This highlights the insensitivity of two `coherent' electronic properties of \Co, heavy-fermion superconductivity and Kondo-lattice screening, to the magnetic configuration of the substituted rare earth ions, the implications of each we will consider in turn.

The pair-breaking effect in unconventional superconductors arises via both potential (non-magnetic) and spin-flip scattering mechanisms. Potential scattering was shown via La substitution in \Co\ to follow the Abrikosov-Gor'kov (AG) model for an anisotropic order parameter \cite{Petrovic_La}, where it is well known that superconductivity is destroyed once the mean free path, $l_{\rm mfp}$, approaches the superconducting coherence length, $\xi$. Here, we estimate this critical scattering length to be $l_{cr} \simeq 180~\AA$ at the point where $T_c\to 0$ (\ie, at  $\rho_{cr}\simeq 20~\mu\Omega$~cm, Fig.~\ref{fig:phase}), assuming that the proportionality between $l_{\rm mfp}(x=0) \simeq 1200~\AA$ \cite{Kasahara} and $\rho(x=0)$ near \Tc\ is independent of doping. This value is roughly twice the in-plane coherence length $\xi_a=80~\AA$ \cite{Petrovic_Co} and consistent with previous work \cite{Petrovic_La}. Interestingly, the value $\rho_{cr}\simeq 20~\mu\Omega$~cm coincides with that found in the series \Sn\ \cite{Bauer_Sn}, where Sn substitution for In preferentially occurs in the Ce-In layers \cite{Daniel_Sn}. In the absence of any dependence on replacement ion size, as evidenced by the contrast in metallic radii of Lu ($1.735~\AA$) and Y ($1.801~\AA$), pair-breaking in \Co\ thus appears to be dominated by general disorder in the CeIn$_3$ planes.

The spin-flip interaction imposed on Cooper pairs by magnetic impurities is characterized by an additional pair-breaking term $\Delta T_c \propto \mathcal{J}^2 D_J$ which includes the exchange interaction parameter $\mathcal{J}$ and the de~Gennes factor $D_J=(g-1)^2J(J+1)$, with the latter reflecting the classic competition between superconductivity and magnetism \cite{dGscaling}. The absence of a dependence of $\Delta T_c$ on this term in \R\ is intriguing, but not unprecedented. In UPt$_3$, the insensitivity of $\Delta T_c$ to $D_J$ is attributable to an odd-parity pairing state, where an equal Zeeman shift on parallel spin states renders the spin-flip process ineffective \cite{Dalichaouch}. In the spin-singlet cuprates, \Tc\ is insensitive to the flavor of the rare earth ion, R, placed in RBa$_2$Cu$_3$O$_{6-\delta}$ \cite{RBCO} owing to the large physical separation between R ions and the CuO$_2$ layers, and hence owing to negligible magnetic interaction. In \Co, evidence for even-parity pairing \cite{Higemoto} also suggests a small value of $\mathcal{J}$, given the drastic range of $D_J$ values (from 0.80 for R=Pr to 15.75 for R=Gd, largest in the rare earth series). However, in contrast to the case of the cuprates, the placement of R ions directly into the active pairing layer \cite{Daniel_Sn} of \Co\ provides the first example of \Tc\ suppression in a spin-singlet superconductor that is truly independent of $D_J$. Assuming the AG model applies, this places stringent bounds on both the strength of the exchange interaction involved in pair-breaking and the nature of the pairing mechanism itself.

\begin{figure}
 \centering
 \includegraphics[totalheight=2.7in]{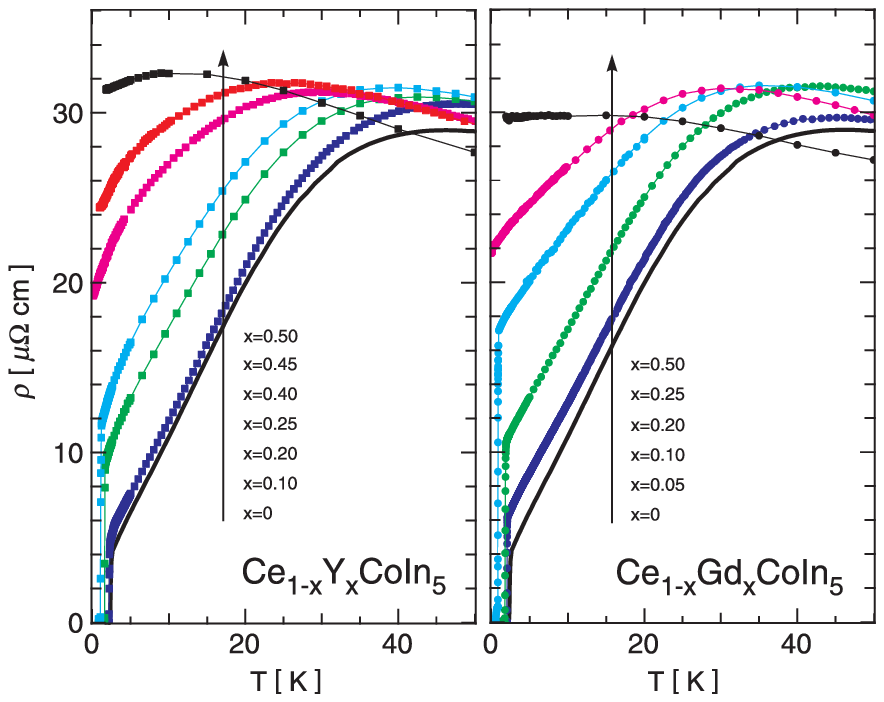}
 \caption{\label{fig:rhoT} 
 {\bf Comparison of electrical resistivity evolution of \R\ with both magnetic and non-magnetic rare-earth substitution.}
 The resistivity $\rho$ is plotted for \Y\ (left) and \Gd\ (right) as a function of nominal concentration of rare-earth substitution. Although both the superconducting transition and Kondo coherence temperature (maximum in $\rho(T)$) are suppressed at the same rate for both substitution series, the temperature dependence of $\rho$ is strongly dependent on the magnetic nature of the substituent ion: Y-doping imposes a strong downward curvature on $\rho(T)$ with increasing concentration, whereas Gd-doping elicits a negligible change in the $T$-linear resistivity present in pure \Co.}
\end{figure}

Interestingly, this insensitivity to $D_J$ is mimicked in the suppression of \Tcoh\ with rare-earth substitution, as shown in Fig.~\ref{fig:phase}. The temperature \Tcoh\ is a characteristic property of the Kondo lattice; associated with the single-ion Kondo temperature $T_K$ \cite{Burdin} and hybridization gap \cite{Dordevic}, it signifies the onset of Kondo singlet formation and marks the scale where single-site magnetic scatterers begin to dissolve into a coherent state. Interestingly, in the same way that superconductivity is destroyed when $l_{\rm mfp} \to l_{cr} \approx \xi$, ~\Tcoh\ also dissappears when $l_{\rm mfp}$ approaches a characteristic coherence length $\xi_{\rm coh} \equiv \hbar v_F/k_B T_{\rm coh} \simeq  100~\AA$ (using $T_{\rm coh} = 50$~K and $v_F \simeq 6.5\times 10^{4}$~m/s, where $k_B$ and$ v_F$ are Boltzmann's constant and the Fermi velocity, respectively) \cite{Cox}, again with no dependence on the magnetism of the dopant ion R. Furthermore, note that $T_{\rm coh} \to 0$ near the $\sim 40 \%$ percolation limit for a 2D lattice. Together these support the notion that, regardless of its internal structure, the Ce lattice vancancy, or `Kondo hole', appears to be the dominant contributor to coherence destruction, leading to a universal dilution of the Kondo lattice as expected by the periodic Anderson model \cite{Wermbter}. Thus, both the superconducting electron pair-breaking effect and the suppression of coherent Kondo screening proceed in a manner that is insensitive to the magnetic configuration of the dopant atom, advancing a scenario where spin-independent disorder is the dominant perturbation in both phenomena.

\begin{figure}
 \centering
 \includegraphics[totalheight=3.3in]{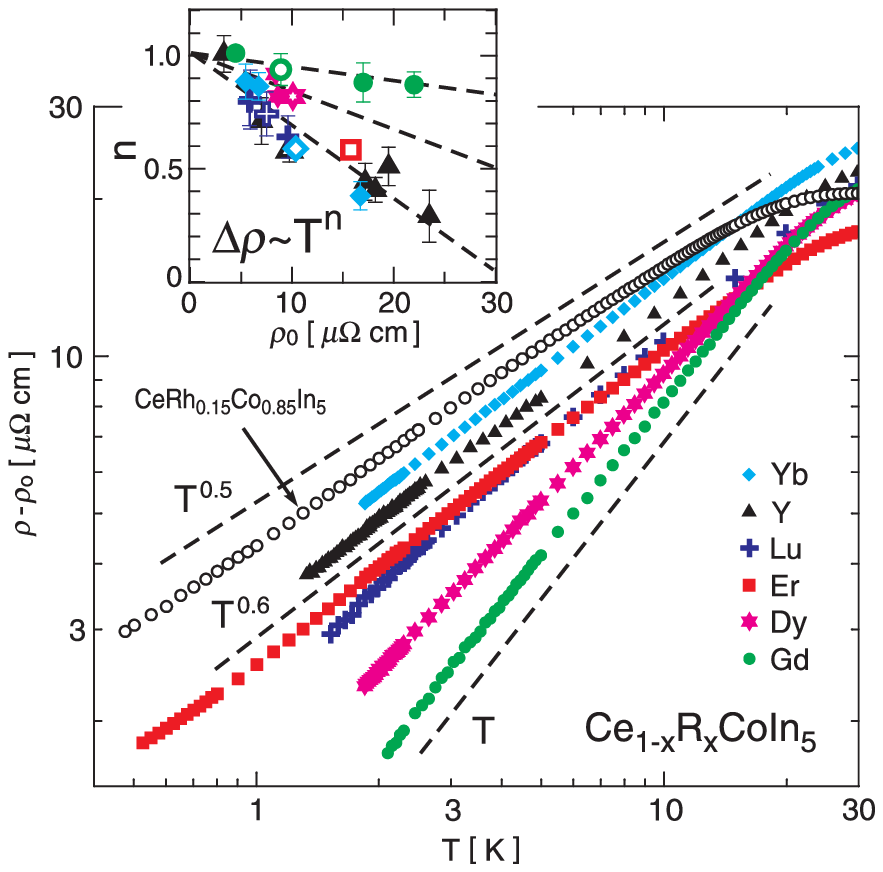}
 \caption{\label{fig:power} 
 {\bf Effect of chemical substitution on $T$-linear resistivity power laws in \Co.} 
 The filled symbols represent various rare-earth substitutions in \R; the open circles (shifted by $\times 2$ for clarity) represent a single-crystal sample of CeRh$_{0.15}$Co$_{0.85}$In$_5$ in its field-induced normal state at 9~T, showing the close connection between sublinear curvature in $\rho(T)$ and the proximity of a spin-density wave instability. The dashed lines are guides exhibiting slopes for various powers of temperature. 
 Inset: Evolution of temperature power-law exponent $n$ (\ie, in $\Delta\rho \sim T^n$) with rare earth substitution (plotted as residual resistivity $\rho_0$ - see Methods section), highlighting the isolated behaviour of Gd substitution. Whereas the large effective moment ($\mu_{\rm eff}=7.9~\mu_B$) of Gd$^{3+}$ ions in \R\ sets it apart from its non-magnetic counterparts, the sublinear exponent observed for Er$^{3+}$ substitution (red square) --  with $\mu_{\rm eff}=9.6~\mu_B$ -- rules out a simple correlation between moment size and sublinear power-law exponent, suggesting the importance of the spin configuration of the rare earth ions and de Gennes factor scaling. The error bars reflect estimates of uncertainty in $n$ based on temperature range and number of data points used in non-linear least squares fits of $\rho(T)$. (The open symbols denote concentrations used in the main figure.)}
\end{figure}

In contrast, the evolution of the non-Fermi liquid electronic transport in \R\ shows a striking sensitivity to the dopant atom's $f$-moment configuration, with $T$-linear resistivity persisting only in the presence of strong local-moment exchange. This is introduced in Fig.~\ref{fig:rhoT} through a direct comparison of the evolution of $\rho(T)$ as a function of both non-magnetic (Y$^{3+}$) and magnetic (Gd$^{3+}$) Ce-site substitution in \R: an increasing Y concentration introduces strong downward curvature in $\rho(T)$ below \Tcoh\ (Fig.~\ref{fig:rhoT}a), whereas $T$-linear scattering seems to be robust against magnetic Gd substitution (Fig.~\ref{fig:rhoT}b).  We further explore this duality by presenting resistivity data for several charateristic rare earth substitutions in Fig.~\ref{fig:power}, fitting $\rho(T)$ for each between \Tc\ and $\sim 20$~K with a simple power law ($\rho = \rho_0 + AT^n$) and plotting $\Delta\rho=\rho-\rho_0$ vs. $T$ to emphasize the exponent $n$, which appears as the slope on a log-log scale. As shown explicity in the inset of Fig.~\ref{fig:power}, $n$ spans a range of sub-linear values, with deviations from $T$-linear being strongest for non-magnetic substitutions.

A sub-$T$-linear transport scattering rate is highly anomalous, yet not unprecedented. For instance, the resistivity of the strongly correlated $f$-electron system Sc$_{1-x}$U$_x$Pd$_3$ was indeed observed to follow the form $\rho(T) = \rho_0 - AT^n$ with an exponent $n\simeq 0.5$ \cite{Dickey}, consistent with the $n=1/2$ expectation of the theoretical multi-channel Kondo model for $T \ll T_K$ \cite{Cox}. However, the $n<1$ curvature in Sc$_{1-x}$U$_x$Pd$_3$ is more likely due to quantum criticality associated with the suppression of spin-glass freezing to $T=0$ near $x_c\simeq 0.3$, rather than the multi-channel Kondo effect \cite{Dickey}. 

Likewise, the phenomenological trend of $n<1$ curvature in \R\ also hints at the proximity of a magnetic instability not unlike that found in \Rh, where similar sublinear curvature is present in $\rho(T)$ above the antiferromagnetic transition at $T_N = 3.8$~K  \cite{Paglione_Rh}. In \Rh, this curvature is proportional to the magnetic entropy, a reflection of the fact that magnetic correlations dominate the transport scattering process \cite{Paglione_Rh}. In \Co\ the same phenomenon was found to be dependent on the proximity to a field-tuned QCP \cite{Paglione_WF}. A connection between the two was established via resistivity measurements of the alloy series \RhCo, where a crossover to sublinear behaviour in $\rho(T)$ was shown to be intimately related to the antiferromagnetic QCP \cite{Jeffries}. As shown in Fig.~\ref{fig:power}, $\rho(T)$ of a single-crystal sample of \RhCo\ with $y=0.85$ (close to the alloy-tuned QCP) indeed follows a $n\simeq 0.5$ exponent over almost two decades in $T$ in its field-induced normal state, indicating a strong connection between $n<1$ scattering and the proximity of a QCP related to the spin-density wave instability in \Rh.

In stark constrast, Gd substitution in \R\ fails to disrupt the mechanism of $T$-linear scattering: 
as shown in the inset of Fig.~\ref{fig:power}, the exponent $n$ experiences an almost negligible change, decreasing at a rate at least five times slower than for non-magnetic substitutions. Because the zero-field magnetic entropy in \Co\ also grows linearly with temperature above \Tc\ \cite{Petrovic_Co}, it is suspected that, like \Rh, magnetic correlations are what shape this anomalous scattering rate. In \Gd, this must involve a Ruderman-Kittel-Kasuya-Yosida (RKKY)-type exchange, as evidenced by both a linear increase with $x$ of the effective moment (up to $\mu_{\rm eff}=7.0~\mu_B$ at $x=1$), and long-range AF order ($T_N\simeq 32$~K at $x=1$) which is in line with the proportionality between $T_N$ and $D_J$ found in other magnetic RCoIn$_5$ compounds \cite{Isikawa}.

But what is the underlying property of Gd$^{3+}$ magnetism that is amenable to $T$-linear scattering? As shown in Fig.~\ref{fig:power}, the curvature in $\rho(T)$ of a sample doped with 25\% Er$^{3+}$ -- with an even larger moment ($\mu_{\rm eff}=9.6~\mu_B$) than Gd$^{3+}$ -- surprisingly exhibits a sublinear power law ($n \simeq 0.6$) much closer to that of the non-magnetic samples. Furthermore, samples doped with Dy$^{3+}$ ($\mu_{\rm eff}=10.6~\mu_B$) exhibit intermediate behaviour, suggesting that the important parameter is not simply moment size itself, but rather involves details of the $f$-moment configuration. In particular, the wide range spanned by the de Gennes factors of Gd$^{3+}$, Dy$^{3+}$ and Er$^{3+}$ (with $D_J$ values of 15.75, 7.08 and 2.55, respectively) is the only aspect of the magnetic configuration that follows the evolution of $n(x)$ suggested by our data set, with a phenomenological form $n \approx 1+ \alpha(D_J - D_0)\rho_0$ where $D_0\simeq 18$ and $\alpha$ is a positive constant. Despite the peculiar position of $D_J$ in the exponent (rather than as a coefficient), its presence highlights the important role of the spin degrees of freedom in the scattering process that gives rise to $T$-linear resistivity, promoting the notion that the `control parameter' may indeed be the projected spin of the scattering centers.

What remains highly anomalous, and more generic, is that the relatively strong relation between $n$ and $D_J$ must comply with the extremely weak exchange coupling between localized $4f$-states and conduction band states, as demonstrated by the insensitivity of both $\Delta T_c$ and $\Delta T_{\rm coh}$ to the magnetic configuration of R. This contrast provides evidence for a separation between the physics of the Kondo lattice and that of the non-Fermi liquid transport in \Co, with the latter necessarily arising from `incoherent' scattering processes. But how can this interaction coexist with the seemingly different long-range interactions that mediate superconductivity and resonant Kondo-lattice screening? One possibility is that the hybridization between $f$-states and conduction-electron states is incomplete, leaving a fraction of incoherent scatterers which conspire to cause such a dichotomy. Evidence for such two-fluid behaviour does indeed take form in \Co, where an `incoherent' fraction of Kondo moments was shown to survive down to \Tc\ \cite{NFP}. Another scenario is of a more profound nature: recent evidence for (1) a group of conduction electrons that remains unpaired in the $T\to 0$ limit \cite{Tanatar_SC} and (2) a direction-dependent violation of the Wiedemann-Franz law \cite{Tanatar_WF} point to a decoupled character of conduction electrons in \Co, suggesting that the separation between the mechanisms behind the coherent properties of \Co\ and its $T$-linear resistivity is of a very fundamental nature.


{\bf Methods:}  Single-crystal platelets of \R\ (including R=Y, Pr, Gd, Dy, Er, Yb and Lu) were grown by the self-flux method \cite{Petrovic_Co}. Samples for measurements of electrical resistivity were prepared with typical dimensions $\sim 2 \times 0.5 \times 0.2$~mm and measured with an a.c. resistance bridge by applying $\sim 0.1$~mA excitation current, directed parallel to the basal plane of the tetragonal crystal structure. The data in Figs.~\ref{fig:phase} and \ref{fig:rhoT} are plotted as a function of residual resistivity in order to eliminate the uncertainty in nominal concentration values. However, note that $\rho_0 \sim x$ to within error as found previously \cite{Petrovic_La,Tanatar_SC}. The d.c. magnetization was measured using a SQUID magnetometer in a 50~mT field, and analyzed using standard Curie-Weiss fits to data between approximately $25$~K and $300$~K to extract effective moments for the magnetic \R\ series. 


{\bf Acknowledgements:} The authors acknowledge B. Coqblin, P. Coleman, C. P\'epin and C. Petrovic for useful discussions and P. Johnson for assistance in sample preparation. Crystal growth and characterization was sponsored by the U.S. Department of Energy (DOE) under
Research Grant DE-FG02-04ER46105, and low-temperature experiments by the National Science Foundation under Grant No. 0335173. J.P. acknowledges support from a NSERC Canada postdoctoral fellowship. Correspondence and requests for materials should be addressed to J.P.


\end{document}